\documentclass[prl,twocolumn,preprintnumbers,amsmath,amssymb]{revtex4}

\usepackage{graphicx}
\usepackage{bm}
\usepackage{color}
\usepackage{ulem}

\begin{document}


\title{Pseudo-Fermi surface and phonon softening in sodium with a stepwise electron distribution}
\author{Shota Ono}
\email{shota\_o@gifu-u.ac.jp}
\author{Daigo Kobayashi}
\affiliation{Department of Electrical, Electronic and Computer Engineering, Gifu University, Gifu 501-1193, Japan}

\begin{abstract}
The absorption of light by a metal disturbs the electron distribution around the Fermi surface. Here, we calculate the phonon dispersion relations of free-electron-like metal, bcc sodium, with a stepwise electron distribution function by using a model pseudo-potential method. The step can behave as a pseudo-Fermi surface, which produces the singularities at specific wavenumbers in the response function. The singularity gives rise to long-range oscillations in the interatomic potential and results in imaginary phonon frequencies around the N point.
\end{abstract}


\maketitle

{\it Introduction.---}The interatomic potential, the sum of the direct interaction energy between the ions and the total electron energy as a function of the ion positions, determines the dynamics of the ions in solids. In his classic textbook \cite{ziman}, Ziman explained, ''In principle, this term would depend on the precise electron configuration, for example, on the number of quasiparticles excited at the temperature of the metal---but in practice it is very insensitive to such difference and usually be computed upon the assumption that the electrons are in their ground state.'' In this paper, we study the effect of the quasiparticle excitations; the deviation from the equilibrium Fermi-Dirac (FD) distribution function at zero temperature.

The finite-temperature density-functional theory (DFT) developed by Mermin \cite{mermin} has enabled us to study the effect of the electron temperature $T_{\rm e}$ on the lattice dynamics of solids. Since the pioneering work of Recoules {\it et al}. \cite{recoules}, the phonon properties that are strongly influenced by the electron thermal excitations have been reported in a variety of simple metals \cite{bottin,giret,minakov,yan,harbour}. Most of the studies have shown that the phonon frequencies of fcc-structured metals increase over the entire Brillouin zone (BZ) when $T_{\rm e}$ is increased up to the Fermi temperature $T_{\rm F}$. In fact, such a phonon hardening has been observed experimentally in gold \cite{ernstorfer}. On the other hand, in bcc-structured metals, a phonon softening appears around the N point within the BZ \cite{giret,yan,harbour}. Similar softening behavior has also been reported in sodium under high pressure \cite{sli}. Recently, we have constructed a model that reproduces the DFT results of phonons in electronically excited aluminium and sodium \cite{ono2019}. 

Experimentally, the ultrafast laser pulse excites the electrons of metals, creating a nonequilibrium distribution function that is usually approximated by a stepwise function \cite{fann,mueller,labouret}. The electron-electron scattering occurs frequently within a subpicosecond timescale, yielding a FD function at $T_{\rm e}$ much larger than the lattice temperature. Although the phonon properties in such a regime have been studied within the finite-temperature DFT \cite{recoules,bottin,giret,minakov,yan,harbour}, those in a nonequilibrium regime has not been investigated so far. 

In this paper, we study the effect of nonequilibrium electron distribution on the phonon dispersion relations of bcc sodium by using a model developed in Ref.~\cite{ono2019}. We model the electron distribution by a stepwise function around the Fermi level. The step can behave as a pseudo-Fermi surface, which produces, in the response function, the singularities at specific wavenumbers that are smaller and larger than twice the Fermi wavenumber. The singularity of the smaller wavenumber gives rise to long-range oscillations in the interatomic potential, which may be referred to as the nonequilibrium Friedel oscillation, and results in imaginary phonon frequencies around the N point. The concept of the pseudo-Fermi surface before reaching the electron quasiequilibrium may be a key to manipulate the lattice dynamics in metals. 


{\it Model.---}A method for calculating the phonon dispersion relation of simple metals under electronic excitations is summarized in Ref.~\cite{ono2019}. Below, we briefly describe a method for treating nonequilibrium electron conditions. 

\begin{figure}[bbb]
\center
\includegraphics[scale=0.5]{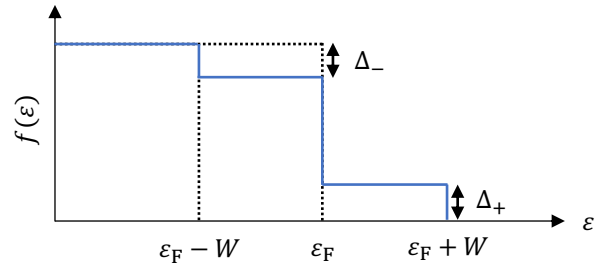}
\caption{\label{fig1} Nonequilibrium electron distribution function defined by Eq.~(\ref{eq:step}). }
\end{figure}

Let us consider a simple metal that consists of ions and conducting electrons, where each ion and electron have charges $Ze$ and $-e$, respectively. With a charge neutrality condition, the number of electrons is uniquely determined when that of ions is given. We express the interatomic potential by 
\begin{eqnarray}
V_{\rm t}(R)
 &=& \frac{Z^2 e^2}{4\pi \varepsilon_0 R}
 + v_{\rm ind}(R),
\label{eq:ion-ion}
\end{eqnarray}
for ions distance $R$ apart. The first term is the direct interaction potential between ions ($\varepsilon_0$ the dielectric constant of vacuum). $v_{\rm ind}$ is the indirect interaction potential that is derived from the electron-mediated ion-ion interaction written as 
\begin{eqnarray}
v_{\rm ind}(R)
 &=& 
 \int_{0}^{\infty} K(q) \frac{\sin (qR)}{qR}dq
\label{eq:ind}
\end{eqnarray}
with the wavenumber $q$. By taking the Hartree, exchange, and correlation energies into account, the kernel $K(q)$ is written as \cite{hartmann,wallace}
\begin{eqnarray}
K(q)
 &=& - \left(\frac{\varepsilon_0 q^4}{2\pi^2 e^2}\right)
 \frac{v_{\rm ps}^{2}(q) v(q)\chi(q)}{1+[1-G(q)]v(q)\chi(q)},
\label{eq:ind2}
\end{eqnarray}
where $v_{\rm ps}(q)$ and $v(q)=e^2/(\varepsilon_0 q^2)$ are the Fourier transform of the model pseudopotential and the bare Coulomb potential, respectively, $\chi(q)$ is the response function, and $G(q)$ is a function for treating exchange and correlation energies. 

As a model, we use the Ashcroft-type potential \cite{ashcroft}, whose Fourier transform is given by
\begin{eqnarray}
 v_{\rm ps} (q) = - Zv(q) \cos(q r_{\rm c}),
 \label{eq:Ashcroft}
\end{eqnarray}
where $r_{\rm c}$ is the cutoff radius. $\chi(q)$ is given by the Lindhard response function
\begin{eqnarray}
\chi(q)
 &=& \frac{m}{\pi^2\hbar^2 q} 
 \int_{0}^{\infty} k f(\varepsilon) 
 \ln \left\vert \frac{2k+q}{2k-q} \right\vert dk,
 \label{eq:response}
\end{eqnarray}
where $\hbar$ is the Planck constant, $m$ is the electron mass, $k$ is the wavenumber, and $f(\varepsilon)$ is the electron distribution function with $\varepsilon = \hbar^2k^2/(2m)$. We use the Hubbard-type function for the exchange and correlation corrections and assume $G(q)=g_1 q^2/(q^2 + g_2)$, where the parameters of $g_1$ and $g_2$ are determined from an analytical formula given in Ref.~\cite{UI}.

Using Eq.~(\ref{eq:ion-ion}) and setting $f(\varepsilon)$ (see below), we diagonalizing the dynamical matrix \cite{ashcroft_mermin} to calculate the phonon frequencies $\omega(\bm{q},\gamma)$, where $\bm{q}$ is the wavevector and $\gamma$ is the branch index, for bcc sodium ($Z=1$). Hereafter, we define the phonon energy as $E_{p}(\bm{q},\gamma) = {\rm sgn}(\omega^2 (\bm{q},\gamma)) \hbar \vert \omega (\bm{q},\gamma)\vert$: $E_{p}$ is negative when $\omega$ is imaginary. The material parameters used are as follows: the lattice constant $a_{\rm lat}=4.225$ \AA, the Fermi energy $\varepsilon_{\rm F}=3.24$ eV, and the cutoff radius $r_{\rm c}=0.906$ \AA \ in Eq.~(\ref{eq:Ashcroft}). The value of $r_{\rm c}$ was optimized in order to reproduce the DFT results at $T_{\rm e}=0$ K \cite{ono2019}. 


\begin{figure}[ttt]
\center
\includegraphics[scale=0.45]{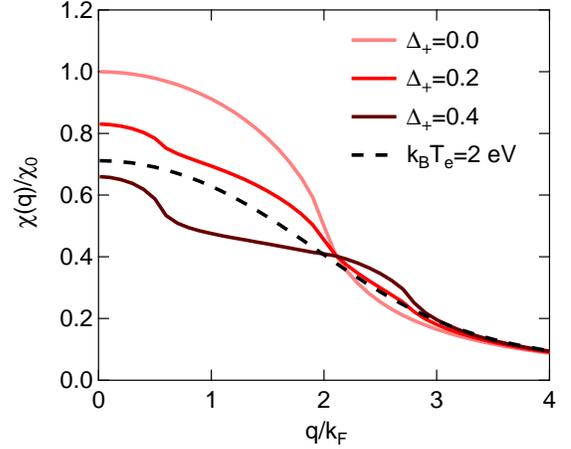}
\caption{\label{fig2} $\chi(q)$ defined by Eq.~(\ref{eq:chiq}) for $\Delta_{+} =0, 0.2,$ and 0.4. The photon energy is $W=3$ eV. The curve of $\chi(q)$ for $k_{\rm B}T_{\rm e}=2$ eV is also shown. $\chi(q)$ is normalized to $\chi_0$ that is equal to $\chi(q=0)$ for $\Delta_+=0$. }
\end{figure}

{\it A stepwise distribution function.---}In our model, the effect of nonequilibrium electron distribution enters into $\chi(q)$ only. The absorption of multiple photons with the energy $W$ disturbs the electron distribution around the Fermi level, resulting in a stepwise function \cite{fann,mueller,labouret}. In the present study, we simply express such a distribution function as 
\begin{eqnarray}
 f(\varepsilon) &=& 
 \theta_{\rm H}((\varepsilon_{\rm F} - W) - \varepsilon )
 \nonumber\\
&+& (1-\Delta_{-}) 
\theta_{\rm H}(\varepsilon - (\varepsilon_{\rm F} - W))
\theta_{\rm H}(\varepsilon_{\rm F} - \varepsilon)
 \nonumber\\
&+& \Delta_{+} 
\theta_{\rm H}(\varepsilon - \varepsilon_{\rm F} )
\theta_{\rm H}((\varepsilon_{\rm F} + W) - \varepsilon),
\label{eq:step}
\end{eqnarray}
where $\theta_{\rm H}$ is the Heaviside step function, and $\Delta_+$ and $\Delta_-$ indicate the deviation from the FD function for $\varepsilon - \varepsilon_{\rm F} > 0$ and $\varepsilon - \varepsilon_{\rm F} < 0$, respectively (see Fig.~\ref{fig1}). Using a charge neutrality condition 
\begin{eqnarray}
 \int_{0}^{\varepsilon_{\rm F}} D(\varepsilon) d\varepsilon
 =  \int_{0}^{\infty} D(\varepsilon) f(\varepsilon)d\varepsilon
\end{eqnarray} 
with the electron density-of-states $D(\varepsilon)\propto \sqrt{\varepsilon}$ (within the free-electron approximation), one obtains the following relation
\begin{eqnarray}
 \frac{\Delta_-}{\Delta_+} = 
 \frac{(\varepsilon_{\rm F} + W)^{3/2} - \varepsilon_{\rm F}^{3/2}}
 {\varepsilon_{\rm F}^{3/2} - (\varepsilon_{\rm F} - W)^{3/2}}.
\end{eqnarray}
Hereafter, we treat $\Delta_+$ as an electron excitation parameter. Analytical expression for $\chi(q)$ under the stepwise distribution of Eq.~(\ref{eq:step}) is given by
\begin{eqnarray}
\chi(q) &=& \frac{m}{\pi^2\hbar^2q} 
\Big[
(1-\Delta_{-} - \Delta_+)k_{\rm F}^2 \alpha_{\rm F} L(\alpha_{\rm F})
\nonumber\\
&+& \Delta_{-} k_{-}^2 \alpha_{-} L(\alpha_{-})
+ \Delta_{+} k_{+}^2 \alpha_{+} L(\alpha_{+})
\Big],
\label{eq:chiq}
\end{eqnarray}
where $\alpha_{j}=q/(2k_j)$ ($j={\rm F}, \pm$) is the dimensionless wavenumber with $k_{\rm F}$ being the Fermi wavenumber and $k_{\pm} = \sqrt{2m(\varepsilon_{\rm F} \pm W)}/\hbar$. We have introduced the function 
\begin{eqnarray}
 L(\alpha) = 1+ \frac{1-\alpha^2}{2\alpha} 
 \ln \left\vert \frac{1+\alpha}{1-\alpha} \right\vert,
\end{eqnarray}
where its first derivative with respect to $q$ diverges at $\alpha = 1$. The profile of $\chi(q)$ for $\Delta_{+} =0, 0.2,$ and 0.4 is shown in Fig.~\ref{fig2}. The value of $W$ is fixed to 3 eV that is nearly equal to the pump energy used in Ref.~\cite{ernstorfer}. As $\Delta_+$ increases, the singularity at $q=2k_{\rm F}$ is smeared out and other singularities appear around $q=2k_{\pm}$ that originate from $L(\alpha_{\pm})$ in Eq.~(\ref{eq:chiq}). No singularities except at $q=2k_{\rm F}$ are observed when the FD function at $k_{\rm B}T_{\rm e}=2$ eV ($k_{\rm B}$ the Boltzmann constant) is assumed (dashed). The singularity around $q=2k_{-}$ will cause a Friedel-like oscillation in $v_{\rm ind}$ for larger $R$, which plays a vital role in explaining the appearance of imaginary frequencies around the N point, as shown below. 

\begin{figure}[ttt]
\center
\includegraphics[scale=0.4]{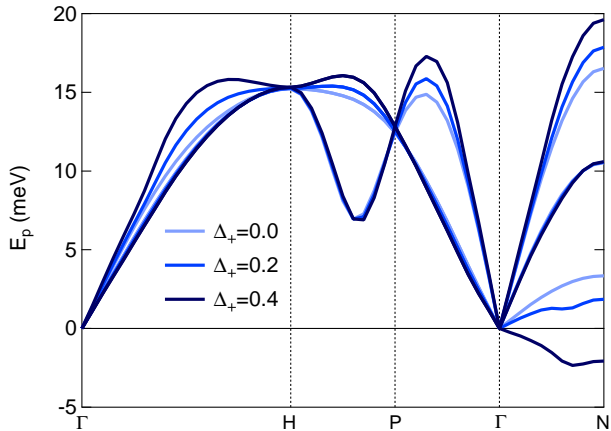}
\caption{\label{fig3} The phonon dispersion relations of sodium along symmetry lines for $\Delta_{+} =0, 0.2,$ and 0.4. The photon energy is fixed to $W=3$ eV.
}
\end{figure}


{\it Results and discussion.---}Figure \ref{fig3} shows the $E_{p}$ along the symmetry lines for sodium for $\Delta_{+} = 0.0, 0.2$, and 0.4. The lowest phonon frequency at the N point decreases significantly with $\Delta_{+}$. When $\Delta_{+}$ is larger than 0.25, imaginary frequencies start to appear around $\bm{q} = \pi/(2a_{\rm lat})$(0,1,1) along the $\Gamma$-N direction. 

\begin{figure}[ttt]
\center
\includegraphics[scale=0.4]{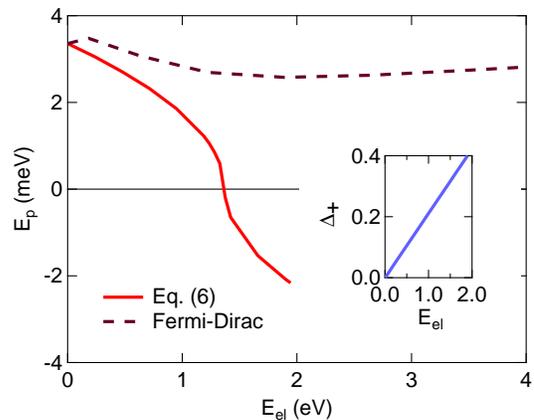}
\caption{\label{fig4} The $E_{\rm el}$-dependence of the lowest $E_{p}$ at the N point under a stepwise distribution of Eq.~(\ref{eq:step}) (solid) and the FD function (dashed). Inset: $\Delta_+$ versus $E_{\rm el}$. }
\end{figure}

The phonon softening behavior above is similar to the previous observations when $T_{\rm e}$ is increased to $T_{\rm F}$ \cite{yan,harbour,ono2019}. To understand how the impact of nonequilibrium electron distribution is different from that of quasiequilibrium distribution with $T_{\rm e}$, we compute the excess electron energy defined as
\begin{eqnarray}
 E_{\rm el} = 2\int_{0}^{\infty} \varepsilon D(\varepsilon) [f(\varepsilon) - f_{\rm FD}(\varepsilon,0)] d\varepsilon,
\end{eqnarray}
where $f_{\rm FD}(\varepsilon,0) = \theta (\varepsilon_{\rm F} - \varepsilon)$ is the FD distribution function at $T_{\rm e}=0$ K. The factor of $2$ comes from the spin degeneracy. A linear relation between $E_{\rm el}$ and $\Delta_+$ holds as shown in the inset of Fig.~\ref{fig4}. Figure \ref{fig4} shows the lowest $E_p$ at the N point as a function of $E_{\rm el}$ for two cases: $f(\varepsilon)$ is equal to Eq.~(\ref{eq:step}) with $\Delta_{+}\ge 0$ and to the FD distribution with $T_{\rm e} \ge 0$ K. For $\Delta_{+}\ge 0$ (solid), $E_p$ decreases with increasing $E_{\rm el}$ and becomes negative when $E_{\rm el}\ge 1.4$ eV/atom. For $T_{\rm e} \ge 0$ K (dashed), on the other hand, no negative values are observed, while $E_p$ takes a minimum value when $E_{\rm el}\simeq 2.0$ eV/atom that corresponds to $k_{\rm B}T_{\rm e}=2$ eV. This comparison means that $E_p$ is not merely a function of $E_{\rm el}$ but a functional of $f(\varepsilon)$. A similar $f(\varepsilon)$-dependence has also been reported in the electron relaxation dynamics \cite{ono2018}.
 
We next elucidate the origin of the imaginary frequency at the N point. The lowest phonon frequency at the N point is formally given by
\begin{eqnarray}
M_{\rm i} \omega_{N}^{2} 
&=&
2\sum_{l=1}^{\infty} \left[
a_l\frac{V_{\rm t}' (R_l)}{R_l}
+ b_l V_{\rm t}'' (R_l)
\right],
 \label{omegaN}
\end{eqnarray}
where $M_{\rm i}$ is the ion mass, $R_l$ ($l=1,2, \cdots$) is the interatomic distance for the $l$th nearest-neighbor sites. The prime of $V_{\rm t}$ means the derivative with respect to $R$ that is evaluated at $R=R_l$. The coefficients $a_l$ and $b_l$ are non-negative values and determined by diagonalizing the dynamical matrix. These values up to $l=8$ are provided in Supplemental Material \cite{suppl}; For example, $a_1=4$, $a_2=2$, $b_1=0$, and $b_2=2$ up to $l=2$. Note that the relation $b_1=0$ holds at the N point for bcc crystals \cite{ono2019}, which results in a significant suppression of the lowest phonon frequency.  

\begin{figure}[ttt]
\center
\includegraphics[scale=0.5]{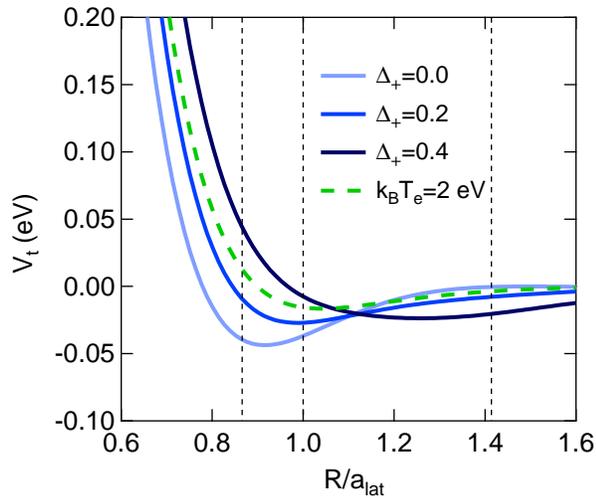}
\caption{\label{fig5} The curve $V_{\rm t}$ given by Eq.~(\ref{eq:ion-ion}) for $\Delta_+ =0, 0.2,$ and 0.4. The case for $k_{\rm B}T_{\rm e}=2$ eV is also shown. The vertical dotted lines indicate the interatomic distance up to the third nearest neighbor sites. }
\end{figure}

Figure \ref{fig5} shows $V_{\rm t}(R)$ for $\Delta_{+} = 0.0, 0.2$, and 0.4. For $\Delta_{+} = 0$, $V_{\rm t}(R)$ is minimum when $R_{\rm min} \simeq 0.91a_{\rm lat}$ that is between $R_1 = \sqrt{3}a_{\rm lat}/2 \ (\simeq 0.866 a_{\rm lat})$ and $R_2 = a_{\rm lat}$. It is enough to consider the terms up to $l=2$ in Eq.~(\ref{omegaN}) because $V_{\rm t}$ is almost flat for larger $R$. It is obvious that $V_{\rm t}' (R_1)$ and $V_{\rm t}' (R_2)$ in Eq.~(\ref{omegaN}) are negative and positive, respectively: They are partially cancelled each other. The positive value of $\omega_{N}^{2}$ is mainly due to the positive $V_{\rm t}''(R_2)$ \cite{ono2019}. On the other hand, when $\Delta_{+} > 0$, the profile of $V_{\rm t}(R)$ changes significantly: The potential minimum shifts toward larger $R$ because $V_{\rm t}$ increases and decreases for smaller and larger $R$, respectively. Both $V_{\rm t}' (R_1)$ and $V_{\rm t}' (R_2)$ become negative, which contribute to the decrease in $\omega_{N}^{2}$. However, it is not enough to explain the appearance of the imaginary frequency at the N point: The terms with, at least, $l\ge 8$ must be included, as listed in Table \ref{table_N}.

\begin{table}[bbb]
\begin{center}
\caption{The lowest $E_{p}$ (in units of meV) at the N point up to the $l$th nearest-neighbor sites considered in Eq.~(\ref{omegaN}). The terms for $l=5$ and $6$ are omitted because $a_l=b_l=0$. $\Delta_{+}=0.4$ is assumed. }
{
\begin{tabular}{cccccccc}\hline
 $l$ \hspace{2mm}& 2 \hspace{2mm}& 3 \hspace{2mm}& 4\hspace{2mm} & 7\hspace{2mm} & 8\hspace{2mm} & $\cdots$ & 120 \\ \hline
 $E_p$ \hspace{2mm}& 2.07\hspace{2mm} & 3.73\hspace{2mm} & 1.45\hspace{2mm} & 0.60\hspace{2mm} & $-1.85$\hspace{2mm} & $\cdots$ & $-2.07$  \\ \hline
\end{tabular}
}
\label{table_N}
\end{center}
\end{table}

\begin{figure}[ttt]
\center
\includegraphics[scale=0.45]{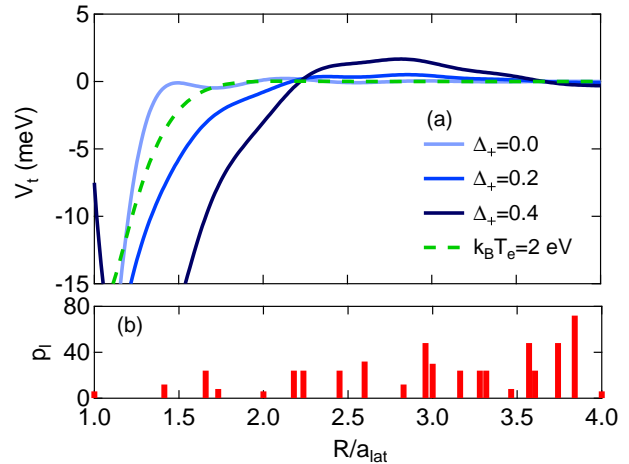}
\caption{\label{fig6} (a) The magnified view of $V_{\rm t}$ for the cases $\Delta_+ \ge 0$ and $k_{\rm B}T_{\rm e}=2$ eV up to $R=4a_{\rm lat}$. (b) The number of the atom with $R=R_l$ ($l=2,\cdots,22$) in a bcc-structured simple metal. }
\end{figure}

To understand such a nature of the long-range interaction, we discuss how the $\Delta_{+}$-dependence of $V_{\rm t}(R)$ is related to the $q$-dependence of $\chi$ shown in Fig.~\ref{fig2}. The impact of $\Delta_{+}$ differs across $q \simeq 2k_{\rm F}$: $\chi(q)$ decreases and increases for smaller and larger $q$s, respectively. This means that the screening of the electron-ion potential is relatively weakened and enhanced for smaller and larger wavenumbers, respectively. In a real-space representation, the electron-ion attractive interaction becomes weakened for smaller $R$, while that becomes enhanced for larger $R$. This scenario, together with Eq.~(\ref{eq:ion-ion}), explains the $\Delta_{+}$-dependence of $V_{\rm t}(R)$ shown in Fig.~\ref{fig5}. In addition, the singularity of $\chi$ at $q=2k_{-}$ causes an enhanced long-range oscillation in $V_{\rm t}$: The wavelength is estimated to be $\pi/k_- \simeq 3 a_{\rm lat}$ from $k_{\rm F}=0.92$ \AA$^{-1}$ and $k_- \simeq k_{\rm F}/4$. Figure \ref{fig6}(a) shows the magnified view of $V_{\rm t}$ up to $R=4a_{\rm lat}$. The number of atoms $p_l$ with $R=R_l$ is also shown in Fig.~\ref{fig6}(b). For $\Delta_{+}=0.4$, the values of $V_{\rm t}'$ and $V_{\rm t}''$ cannot be negligible even for $R\ge 1.5a_{\rm lat}$ as mentioned, and $V_{\rm t}''$ is negatively large: For example, $V_{\rm t}''(R_4) = -6.9$ and $V_{\rm t}''(R_8) = -2.7$ meV/\AA$^2$ with $R_4=\sqrt{11}a_{\rm lat}/2$ and $R_8=\sqrt{5}a_{\rm lat}$. From Eq.~(\ref{omegaN}), these $V_{\rm t}''$ can contribute to the decrease in the lowest $E_p$ at the N point, yielding the imaginary frequency. On the other hand, $V_{\rm t}$ almost decays at $R\simeq 1.5a_{\rm lat}$ for the cases of $\Delta_{+}=0$ and $k_{\rm B}T_{\rm e}=2$ eV because no singularities below $q=2k_{\rm F}$ are present, as shown in Fig.~\ref{fig2}.

{\it Conclusion.---}We have studied the effect of nonequilibrium electron distribution on the phonon dispersion relations for bcc sodium. The nonequilibrium distribution is assumed to be a stepwise function around the Fermi level. We have shown that the step-induced singularity at specific wavenumbers in the response function causes an anomalously enhanced Freidel-like oscillation of the interatomic potential, giving rise to imaginary phonon frequencies around the N point. We expect that a pseudo-Fermi surface created by the photon absorption may be a key to manipulate the lattice dynamics of metals before the electron thermalization. 

In the present study, the impact of the nonequilibrium distribution $f(\varepsilon)$ was considered only: The ions are fixed to the equilibrium position at $T_{\rm e}=0$ K by assuming isochoric heating. The ion displacements in electron excited metals would lead to solid-to-solid transformation \cite{giret}, melting \cite{ernstorfer}, and ablation \cite{rethfeld2017}. Such studies are left for future work. 

 


\begin{thebibliography}{99}

\bibitem{ziman} J. M. Ziman, {\it Principles of the Theory of Solids} (Cambridge University Press, Cambridge, 1972).

\bibitem{mermin} N. D. Mermin, Thermal properties of the inhomogeneous electron gas, Phys. Rev. {\bf 137}, A1441 (1965).

\bibitem{recoules} V. Recoules, J. Cl\'{e}rouin, G. Z\'{e}rah, P. M. Anglade, and S. Mazevet, Effect of Intense Laser Irradiation on the Lattice Stability of Semiconductors and Metals, Phys. Rev. Lett. {\bf 96}, 055503 (2006).

\bibitem{bottin} F. Bottin and G. Z\'{e}rah, Formation enthalpies of monovacancies in aluminum and gold under the condition of intense laser irradiation, Phys. Rev. B {\bf 75}, 174114 (2007).

\bibitem{giret} Y. Giret, S. L. Daraszewicz, D. M. Duffy, A. L. Shluger, and K. Tanimura, Nonthermal solid-to-solid phase transitions in tungsten, Phys. Rev. B {\bf 90}, 094103 (2014).

\bibitem{minakov} D. V. Minakov and P. R. Levashov, Melting curves of metals with excited electrons in the quasiharmonic approximation, Phys. Rev. B {\bf 92}, 224102 (2015).

\bibitem{yan} G. Q. Yan, X. L. Cheng, H. Zhang, Z. Y. Zhu, and D. H. Ren, Different effects of electronic excitation on metals and semiconductors, Phys. Rev. B {\bf 93}, 214302 (2016).

\bibitem{harbour} L. Harbour, M. W. C. Dharma-wardana, D. D. Klug, and L. J. Lewis, Equation of state, phonons, and lattice stability of ultrafast warm dense matter, Phys. Rev. E {\bf 95}, 043201 (2017).

\bibitem{ernstorfer} R. Ernstorfer, M. Harb, C. T. Hebeisen, G. Sciaini, T. Dartigalongue, R. J. D. Miller, The Formation of Warm Dense Matter: Experimental Evidence for Electronic Bond Hardening in Gold, Science {\bf 323}, 1033 (2009).

\bibitem{sli} S. Li, C. Wang and Y. Chen, Anharmonic lattice dynamics of bcc sodium under high pressures, Phys. Chem. Chem. Phys. {\bf 20}, 14647 (2018).

\bibitem{ono2019} S. Ono, Lattice dynamics for isochorically heated metals: A model study, J. Appl. Phys. {\bf 126}, 075113 (2019).


\bibitem{fann} W. S. Fann, R. Storz, H. W. K. Tom, and J. Bokor, Electron thermalization in gold, Phys. Rev. B {\bf 46}, 13592 (1992). 

\bibitem{mueller} B. Y. Mueller and B. Rethfeld, Relaxation dynamics in laserexcited metals under nonequilibrium conditions, Phys. Rev. B {\bf 87}, 035139 (2013).

\bibitem{labouret} T. Labouret and B. Palpant, Nonthermal model for ultrafast laser-induced plasma generation around a plasmonic nanorod, Phys. Rev. B {\bf 94}, 245426 (2016). 

 
 


\bibitem{hartmann} W. M. Hartmann and T. O. Milbrodt, Model-Potential Calculations of Phonon Energies in Aluminum, Phys. Rev. B {\bf 3}, 4133 (1971).

\bibitem{wallace} D. C. Wallace, {\it Thermodynamics of crystals} (New York, Wiley, 1972).

\bibitem{ashcroft} N. W. Ashcroft, Phys. Lett. {\bf 23}, 48 (1966).


\bibitem{UI} S. Ichimaru and K. Utsumi, Analytic expression for the dielectric screening function of strongly coupled electron liquids at metallic and lower densities, Phys. Rev. B {\bf 24}, 7385 (1981).

\bibitem{ashcroft_mermin} N.W. Ashcroft, N. D.Mermin, and D.Wei, {\it Solid State Physics}, revised edition, (Cengage, Boston, 2016).

\bibitem{ono2018} S. Ono, Thermalization in simple metals: Role of electron-phonon and phonon-phonon scattering, Phys. Rev. B {\bf 97}, 054310 (2018).

\bibitem{suppl} See Supplemental Material at XXX for the analytical expression for the dynamical matrix at the N point. 

\bibitem{rethfeld2017} B. Rethfeld, D. S. Ivanov, M. E. Garcia and S. I. Anisimov, Modelling ultrafast laser ablation, J. Phys. D: Appl. Phys. {\bf 50}, 193001 (2017).










\end{thebibliography}
\end{document}